\documentclass[%
 reprint,
 amsmath,amssymb,
 aps,
]{revtex4-2}
\usepackage{amsthm}

\usepackage{import}
\usepackage{physics}
\newtheorem{theorem}{Theorem}
\usepackage[utf8]{inputenc}
\usepackage{lineno,hyperref}
\usepackage{mathtools}
\usepackage{siunitx}
\usepackage{graphicx}
\usepackage{tabularx}
\usepackage{booktabs}

\newtheorem{lemma}[theorem]{Lemma}
\newtheorem{prop}{Proposition}
\newtheorem{definition}{Definition}[section]
\usepackage{dcolumn}
\usepackage{bm}
\usepackage{bbm} 
\usepackage{xcolor}
\newcommand{\be}{\begin{equation}}
\newcommand{\ee}{\end{equation}}
\newcommand{\bq}{\begin{eqnarray}}
\newcommand{\eq}{\end{eqnarray}}
\newcommand{\bea}{\begin{eqnarray}}
\newcommand{\eea}{\end{eqnarray}}
\newcommand{\ba}{\begin{align}}
\newcommand{\ea}{\end{align}}

\begin{document}

\title{Enhancing Quantum Machine Learning: The Power of Non-Linear Optical Reproducing Kernels}

\author{Shahram Dehdashti$^{1,*}$}
\email{shahram.dehdashti@tum.de}
\author{Prayag Tiwari$^{2,*}$}
\email{prayag.tiwari@hh.se}
\author{ Kareem H. El Safty$^{1,*}$}
\email{kareem.el-safty@tum.de}
\author{Peter Bruza$^{3}$}
\email{p.bruza@qut.edu.au}
\author{Janis N\"{o}tzel$^{1}$}
\email{janis.noetzel@tum.de}
\affiliation{$^{1}$Emmy-Noether Gruppe Theoretisches Quantensystemdesign Lehrstuhl F\"{u}r Theoretische Informationstechnik
Technische Universit\"{a}t M\"{u}nchen.}
\affiliation{$^{2}$School of Information Technology, Halmstad University, Sweden}
\affiliation{$^{3}$School of Information Systems, Queensland University of Technology, Australia}
\thanks{S.D, P.T and K. H. S  contributed equally and share co-first authorship}



\begin{abstract}
Amidst the array of quantum machine learning  algorithms, the quantum kernel method has emerged as a focal point, primarily owing to its compatibility with noisy intermediate-scale quantum devices and its promise to achieve quantum advantage. This method operates by nonlinearly transforming data into feature space constructed with quantum states, enabling classification and regression tasks.  In this study, we present a novel feature space constructed using Kerr coherent states, which generalize $su(2)$, $su(1,1)$ coherent states, and squeezed states. Notably, the feature space exhibits constant curvature, comprising both spherical and hyperbolic geometries, depending on the sign of the Kerr parameter. Remarkably, the physical parameters associated with the coherent states,
enable control over the curvature of the feature space.
Our study employs Kerr kernels derived from encoding data into the phase and amplitude of Kerr coherent states. We analyze various datasets ranging from Moon to breast cancer diagnostics. Our findings demonstrate the robustness of Kerr coherent states, attributed to their flexibility in accommodating different hyperparameters, thereby offering superior performance across noisy datasets and hardware setups.
\end{abstract}

\maketitle


\section{Introduction}
Quantum machine learning seeks to leverage the unique capabilities of quantum computing to enhance the performance of machine learning models, either by increasing their speed or accuracy \cite{biamonte2017quantum}. In the early stages of research, efforts were directed towards achieving substantial speed improvements using the HHL algorithm for matrix inversion \cite{harrow2009quantum}, which serves as a fundamental component for various algorithms like quantum support vector machines (SVM) \cite{rebentrost2014quantum} and quantum principal component analysis (PCA) \cite{lloyd2014quantum}. Additionally, researchers explored the identification of topological properties within data \cite{lloyd2016quantum}.
However, these approaches had certain prerequisites, including access to reliable quantum hardware, stringent assumptions regarding data (such as data sparsity), and the reliance on QRAM \cite{giovannetti2008quantum}, an envisioned hardware device designed for efficient quantum state access. More recent research has shifted its focus towards developing algorithms suitable for Noisy Intermediate-Scale Quantum (NISQ) devices, which are quantum computers with limited qubits and some level of noise \cite{preskill2018quantum}. Prominent techniques in this context include variational quantum algorithms like Variational Quantum Eigensolver (VQE) \cite{peters2022generalization} and Quantum Approximate Optimization Algorithm (QAOA) \cite{farhi2014quantum,zhou2020quantum}, which combine both quantum and classical hardware to tackle problems related to materials science and combinatorial optimization.
Furthermore, Quantum Neural Networks have emerged as an approach for addressing supervised learning tasks \cite{abbas2021power}. Among these innovative techniques, quantum kernels stand out as particularly promising \cite{havlicek2019supervised}. 

In fact, quantum kernels represent potent data analysis techniques as they enable the creation of non-linear predictors by converting classical data into a quantum framework. In the realm of machine learning, classical kernel methods have found widespread applications across various scientific domains, including biology, chemistry, and physics \cite{shawe2004kernel, hofmann2008kernel, scholkopf2004kernel, ralaivola2005graph, karniadakis2021physics}. Notably, many kernel method algorithms, such as Ridge Regression and kernel Principal Component Analysis (KPCA), rely on convex optimization and can be efficiently solved using classical computing hardware. Kernel methods employ symmetric functions that are positive definite to map data points into a Reproducible Kernel Hilbert Space (RKHS), by which any function can be recovered \cite{perelomov1977generalized}. 

Quantum kernels are based on the
insight that if we encode a data input $\mathbf{x} \in \mathbb{R}^{n}$ into a quantum
state $\rho(\mathbf{x})=\hat{U}(\mathbf{x})\ketbra{0}{0}\hat{U}^{\dagger}(\mathbf{x})$, e.g., via a quantum state preparation,
we can draw a comparison  $\rho(\mathbf{x})$ with a state $\rho(\mathbf{w})$ via $f(x)=\Tr\left[\hat{\rho}_{w}\hat{\rho}_{x}\right]$. The idea can be understood as a supervised learning task in machine learning when $\hat{\rho}(w)$ is trainable. The radial basis function (RBF) can be considered a well-known example in machine learning. This function is characterized by its definition as $K(x, w) = \exp(-|x - w|^{2} / 2\sigma^{2})$.
In this expression, $x$ and 
$w$ represent two distinct 
sample elements, and the parameter 
$\sigma$ plays a crucial role in 
determining the decision boundary 
\cite{orr1996introduction,musavi1992neural,buhmann2000acta}. It's important 
to highlight that the RBF can also be constructed 
by the inner product of two coherent states, i.e., $\braket{x}
{w}$\cite{kubler2019phys}, 
by considering $\hat{U}(x)$ as a 
displacement operator $\hat{D}
(x)=\exp\left(x\hat{a}^{\dagger}-
x\hat{a}\right)$, where 
$\hat{a}^{\dagger}$ and $\hat{a}$ are 
the creation and annihilation 
operators, respectively.

Coherent states 
according to 
the definition,  
fulfill the resolution of the 
identity $\int d \mu(\alpha) 
\ketbra{\alpha}{\alpha}=\mathbb{I}$, $\alpha\in \mathbb{C}$ 
which leads to the fact that any arbitrary state $\ket{\psi}$ is a  linear combination of coherent states,  $\int d\mu(\alpha) \ket{\alpha}\braket{\alpha}{\psi}=\ket{\psi} $ \cite{perelomov1977generalized}. 
Defining  the function $\psi(\alpha^{\ast}) \equiv \braket{\alpha}{\psi}$ for a coherent state $\ket{\alpha}$ implies the one-to-one correspondence between the states $\ket{\psi}$  and the functions $\psi(\alpha)$ in the complex $\alpha$ plane. Thus, an immediate consequence of the latter is the reproducing kernel property, i.e.,  
\begin{equation*}
    \psi(\beta) = \int d\mu(\alpha) \, K(\beta, \alpha) \psi(\alpha).
\end{equation*}
in which  a kernel is defined as $K(\alpha, \beta) = \langle \alpha | \beta \rangle$.

As previously discussed, 
there are various 
proposals for mapping data into coherent and squeezed states, i.e., phase and amplitude mapping \cite{schuld2019quantum,kubler2019phys,shahram1}. 
Mapping the data into the phase of squeezed states holds the potential for improved outcomes, even though it's noteworthy that the squeezed kernel has no hyperparameter \cite{schuld2019quantum}.

In this paper, we introduce Kerr coherent states (KCSs), which are generalizations of $su(2)$ and $su(1,1)$ coherent states, and realize squeezed states for specific values of the Kerr parameter. 
However, in the case of the Kerr kernel, the Kerr parameter, which serves as the "bandwidth" parameter, and the Hilbert space's dimension control the decision boundary, see  Appendix \ref{sec.app.b}, for detail.  
Moreover, our study reveals 
 Kerr coherent states, geometrically,   correspond to a hyperbolic space for positive Kerr parameters and a spherical space for negative Kerr parameters. In this context, the Ricci scalar is related to the physical parameters of the Kerr coherent states. It is a widely recognized fact that the mapping of data into curved and pseudo-Riemannian manifolds provides a more effective characterization of learned latent spaces for machine learning compared to using a Euclidean space \cite{Arvanitidis2018}. For this reason, various approaches such as Lie Group Machine Learning \cite{Lu2020}, manifold optimization \cite{Boumal2023}, and Machine Learning in hyperbolic spaces have been proposed and recommended \cite{Prakash2023,Hong2023,Qu2022,Fang2023,Mettes2023,Gao2022,Atigh2022}. Specifically, focusing on the Hyperbolic space is of significance \cite{Yang2024,Yang,Mauran2023,Fang2021}. This concept, in turn, opens up possibilities for quantum machine learning on manifolds.

In machine learning, supervised learning involves training an algorithm with pairs of data and corresponding labels $(\mathbf{x}, y) \in \chi \times \{0, 1\}$. 
One of the fundamental models is that of binary classification, where the data is given by a series $\mathbf x$ of data-points $ x_i$ which are all taken from a set $\chi$, each coming with a binary label $y$. If not explicitly states otherwise, we will in the following assume that $\chi=\mathbb R^N$ for some $N\in\mathbb N$. 
The goal is to learn a function $f: \chi \rightarrow \{0, 1\}$ that accurately predicts labels for both training and test data. One popular algorithm for supervised learning is the Support Vector Machine (SVM), which learns a robust linear classification boundary in the input space by using \emph{Euclidean} inner products $\braket{x_{i}}{x_{j}}$ between data points \cite{Cortes1995,Boser1992}.
As mentioned, when handling non-linearly separable data, SVM classifiers can be extended using the kernel trick. In this approach, the data is mapped into a new space via a Kernel function denoted as $\Phi: \mathbf{x} \rightarrow \Phi(\mathbf{x})$, and the inner products are computed as $K_{ij} = \braket{\Phi(x_{i})}{\Phi(x_{j})}$. Each kernel function corresponds to an inner product on input data mapped into a feature Hilbert space $\Phi(\chi)$. The linear classification boundaries established by an SVM trained on this feature space may exist in a higher dimension or on a different manifold with the same dimension \cite{Aizerman1964,Aronszajn1950}. In this context, the Kerr Kernel is introduced, delineating that the feature space manifests as either a pseudo-sphere or a sphere contingent upon the sign of the Kerr parameter. 

Determining whether a specific kernel is well-suited for a dataset can be challenging without advanced knowledge of the data generation process. However, empirical evidence shows that certain families of classically challenging kernels could lead to performance improvements in practice. By introducing different kernels obtained by the Kerr coherent states, we examine different datasets to evaluate the kernels in the SVM. We utilize SVM with various kernels derived from Kerr coherent states across diverse datasets, encompassing the Moon, Circles, Periodic datasets, and the Breast Cancer dataset. Our findings highlight that the ratio of the frequency of light to the Kerr parameter plays a pivotal role in achieving optimal performance. Notably, this parameter directly determines the curvature of the feature space surface, underscoring its significance in the classification process. 
In other aspect, 
by design, our approach flexibly interpolates between "quantum" and "classical" regimes, solely via the control of physical parameters. We therefore believe it to be of interest in the sense on analog computation. Moreover, it highlights the potential of physics-inspired programming, where physical interactions are controlled and programmed to aid in specific computational tasks.

\section{Theoretical aspects of Kerr Coherent state's Kernels}\label{sec_II}
%
In the following, we introduce the Kerr coherent states as the feature space and study their mathematical characters.
\subsection{Kerr Feature Space}
A Kerr coherent state is defined as 
\begin{eqnarray}
\ket{\alpha;\lambda, j}=e^{\alpha \hat{A}^{\dagger}-\alpha^{\ast}\hat{A}}\ket{0},\ \alpha=re^{i\phi} \in \mathbb{C}
\end{eqnarray}
The Kerr annihilation operators  are respectively defined by  $\hat{A}=\sqrt{\frac{\lambda}{2}}\hat{a}\sqrt{2j-1+ \hat{n}}$ and $\hat{A}=\sqrt{\frac{|\lambda|}{2}}\hat{a}\sqrt{2j+1- \hat{n}}$, for positive and negative sign of the Kerr parameter $\lambda \in \mathbb{R}$; $\hat{a}$ and $\hat{a}^{\dag}$ are annihilation and  creation operator, and  $\hat{n}=\hat{a}^{\dagger}\hat{a}$ is the number operator; the Kerr creation operator $\hat{A}^{\dagger}$  is the conjugate transpose of the Kerr annihilation operator $\hat{A}$. For the positive value of the Kerr parameter,  the Kerr
coherent state can be expanded in terms of the Fock state basis 
as 
\begin{eqnarray}
\ket{\alpha;\lambda^{+},j}&=&\cosh^{-2j}\left[\sqrt{\frac{\lambda}{2}} |\alpha| \right] \\
&\times&\sum_{n=0}^{\infty} \sqrt{\frac{\Gamma(2j+n)}{\Gamma(2j) n!}} e^{-i n\phi}\tanh^{n}\left[\sqrt{\frac{\lambda}{2}} |\alpha| \right] \ket{n},\nonumber
\end{eqnarray}
where $j$ can be an integer or half-integer \cite{Dehdashti2022}. It's important to note that when $\lambda$ is set to 2, the $su(1,1)$-coherent state  is obtained. Furthermore, when $\lambda$ is 2 and $j$ is $1/2$, the result is a squeezed state.

For negative values of the Kerr parameter $\lambda$,
the Kerr coherent state $\ket{\alpha;\lambda^{-},j}$
is given by 
\begin{eqnarray}
\ket{\alpha;\lambda^{-},j}&=&\cos^{2j}\left[\sqrt{\frac{|\lambda|}{2}} |\alpha| \right] \\
&\times&\sum_{n=0}^{2j} \sqrt{\frac{(2j)!}{(2j-n)!n!}} e^{-in\phi}\tan^{n}\left[\sqrt{\frac{|\lambda|}{2}} |\alpha| \right] \ket{n},\nonumber
\end{eqnarray}
Note that for the $\lambda=-2$, the above Kerr coherent state represents a $su(2)$-coherent state. 
\begin{figure}
    \centering
\includegraphics[width=7cm]{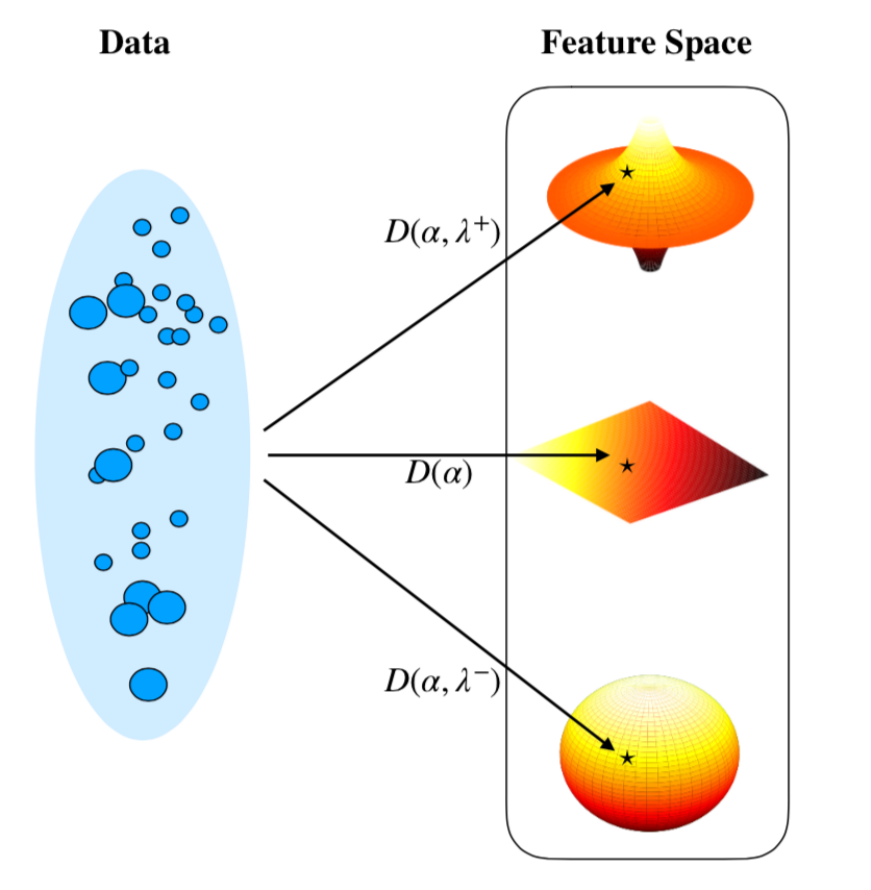}
    \caption{Data by using the displacement operator is transferred in the feature space, which can be pseudosphere, flat and sphere according to the sign of the Kerr parameter. }
    \label{fig_HFS}
\end{figure}

It's important to note that even though the Hilbert space representation of the Kerr coherent states, whether in the negative or positive case, involves more than two dimensions, these states exist on a surface with distinct topological characteristics that depend on the sign of the Kerr parameter. In fact, the feature space created by a positive Kerr coherent state forms a pseudosphere with consistently negative curvature as Fig \ref{fig_HFS} schematically illustrates. For mathematical detail see Appendix \ref{appenddix_A}.

The inner product between two Kerr coherent states, for $\lambda \in \mathbb{R}^{+}$, is given by
\begin{eqnarray}\label{eq_inner_pos}
    K(\alpha_{1},\alpha_{2})=
\frac{\left(\sech^{2}\left[\sqrt{\frac{\lambda}{2}}r_{1} \right]\sech^{2}\left[\sqrt{\frac{\lambda}{2}}r_{2}\right]\right)^{j}}{\left[1-e^{i(\phi_{1}-\phi_{2})}\tanh\left(\sqrt{\frac{\lambda}{2}}r_{1}\right)\tanh\left(\sqrt{\frac{\lambda}{2}}r_{2}\right)\right]^{2j}}.\nonumber\\ 
\end{eqnarray}
where $\alpha_{j}=r_{j}\exp(i\phi_{j})$, with $j=1,2$. For $\lambda \in \mathbb{R}^{-}$, the inner product is given by
\begin{eqnarray}\label{eq_inner_neg}
K(\alpha_{1},\alpha_{2})=\frac{\left[1+e^{i(\phi_{1}-\phi_{2})}\tan\left[\sqrt{\frac{|\lambda|}{2}}r_{1}\right]\tan\left[\sqrt{\frac{|\lambda|}{2}}r_{2}\right]\right]^{2j}}{\left(\sec^{2}\left[\sqrt{\frac{|\lambda|}{2}}r\right]\sec^{2}\left[\sqrt{\frac{|\lambda|}{2}}r^{\prime}\right]\right)^{j}}.\nonumber\\
\end{eqnarray}
When it comes to encoding data within quantum states, we have two distinct methods at our disposal. The first method involves encoding the data into the phase of the quantum states, which is commonly referred to as phase encoding. Alternatively, the second method allows us to map the data to the amplitude of the quantum state, known as amplitude encoding. In the following, we consider both cases in detail.\\

\noindent {\bf Phase encoding:}
In the context of phase encoding, we can take an N-dimensional dataset  $\{x_{m}\}$ and represent the data by utilizing the phase attributes of $N$ Kerr coherent states. The specific choices for the amplitude $|\alpha|=c$, Kerr parameter $\lambda$, and $j$ defined as integer and half-integer  serve as hyperparameters for this encoding process, i.e., {}$K(x_{m},x_{m^{\prime}})=\braket{c,x_{m};\lambda}{c,x_{m^{\prime}};\lambda}$ using relations (\ref{eq_inner_pos}) and (\ref{eq_inner_neg}). We can call it the Kerr phase
kernel. It's worth highlighting that the Kerr phase kernel can be considered an extension of the well-known squeezed phase kernel, a widely employed concept in quantum machine learning \cite{Schuld2021,Li2021} and exhibits analogous characteristics  to the exponential sine squared (ESS) kernel commonly encountered in machine learning:
\begin{eqnarray}\label{eqcss}
    K_{ess}(x_{m}-x_{m^{\prime}})=\exp\left[-\frac{2}{l^{2}}\sin^{2}\left(\frac{\pi}{p}|x_{m}-x_{m^{\prime}}|\right)\right]
\end{eqnarray}
where $l$ and  $p$ are hyperparameters of the length scale and
the period, respectively \cite{Li2021}. 
\begin{figure*}
\includegraphics[width=18cm]{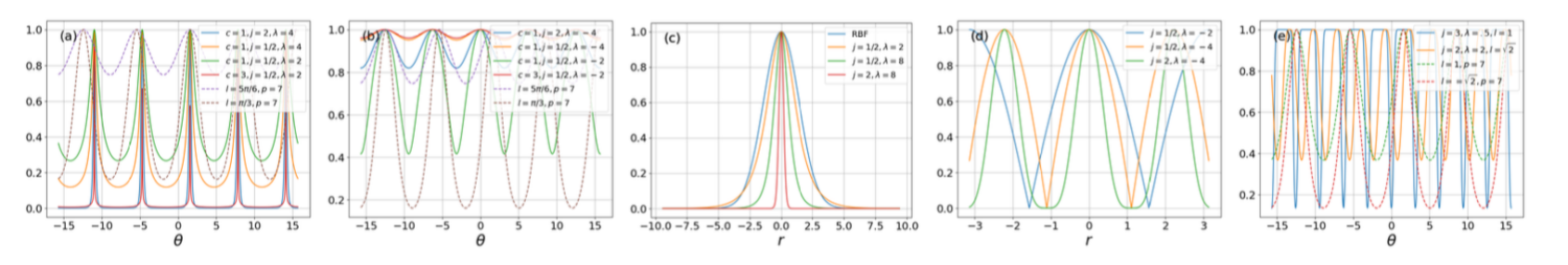}
\caption{The Kerr phase kernel with positive Kerr parameters, represented by Eq. (\ref{eq_inner_pos}, is depicted in the plot (a) across various parameter values and compared with the exponential sine-squared kernel (Eq. (\ref{eqcss})). Plot (b) illustrates the Kerr phase kernel with negative Kerr parameters, i.e., Eq. (\ref{eq_inner_neg}). Plots (c) and (d) showcase the positive and negative amplitude Kernels, specifically Eqs. (\ref{eq_pos_amp}) and (\ref{eq_neg_amp}), respectively. Finally, plot (e) demonstrates the exponential sine-squared kernel (Eq. (\ref{eqcss})) and the exponential cosine kernel (Eq. (\ref{qecs})) for different parameters.}\label{boundry_kernels}
\end{figure*}
These kinds of kernels are suitable for data that has features of periodicity, such as weather forecasting data and the stock market
data.

\noindent {\bf Amplitude encoding:}
Another way to encode into the Kerr coherent state is to encode the data into the amplitude, i.e., $\phi: x_{m}\rightarrow \ket{x_{m},\theta;\lambda}$. In such case, $\theta$, $\lambda$ and $j$ can be hyperparameters. The direct calculation shows that whenever we set $\theta=0$, we can write the Kernel for the positive Kerr parameters as follows:
\begin{eqnarray}\label{eq_pos_amp}
    K(x_{m},x_{m^{\prime}})=\cosh^{-2j}\left[\sqrt{\frac{\lambda}{2}}|x_{m}-x_{m^{\prime}}|\right], 
\end{eqnarray}
which is generalization of the squeezed amplitude kernel, that is $K_{sa}=\cosh^{-1/2}\left[|x_{m}-x_{m^{\prime}}|\right]$.  
For the negative Kerr parameters, the Kernel is given by
\begin{eqnarray}\label{eq_neg_amp}
    K(x_{m},x_{m^{\prime}})=\cos^{2j}\left[\sqrt{\frac{\lambda}{2}}|x_{m}-x_{m^{\prime}}|\right]. 
\end{eqnarray}
It's worth noting that the positive Kerr kernel represents a broader concept encompassing the hyperbolic kernels \cite{Vedaldi2012,Wilson2014}, while the negative Kerr kernel extends the idea of the squared cosine kernel \cite{Schuld2023,Fang2021}.

In addition, following the construction principles of Kernels \cite{hofmann2008kernel}, we propose the corresponding quantum counterpart of the aforementioned Kernel (\ref{eqcss}) as follows:
\begin{eqnarray}\label{qecs}
K_{qecs}(x_{m}-x_{m^{\prime}})=\exp
\left[ -\frac{2}{l^{2}}
\cos^{2j}\left(\sqrt{\frac{|\lambda|}{2}} |x_{m}-x_{m^{\prime}} | \right)
    \right]\nonumber\\
\end{eqnarray}
which is called quantum exponential cosine Kernel (QEC).
Figure \ref{boundry_kernels} presents a comparative analysis of the Kerr phase kernel, Kerr amplitude kernels with varying parameters $j$ and $\lambda$, and exponential sine-squared and exponential cosine kernels. Upon comparison with the squeezing phase, RBF, and exponential sine-squared kernels, it is evident that the hyperparameters of the Kerr phase kernel offer greater control over the distribution boundaries. This heightened flexibility renders the Kerr phase kernel a more versatile option for data classification.

%

\begin{table}[!hbt]
\centering
\caption{Datasets description}
\label{dataset_details}
\begin{tabular}{@{}ccccc@{}}
\toprule
\textbf{Dataset} & \textbf{\# Train} & \textbf{\# Test} & \textbf{Noise} & \textbf{Noisy labels} \\ \midrule
Moons v1     & 300 & 100 & 0.25    & -  \\ \midrule
Moons v2     & 645 & 215 & 0.25    & -  \\ \midrule
Circles v1   & 300 & 100 & 0.1/0.8 & -  \\ \midrule
Circles v2   & 645 & 215 & 0.1/0.8 & -  \\ \midrule
Hypercube v1 & 300 & 100 & 8/4/8   & 40 \\ \midrule
Hypercube v2 & 645 & 215 & 8/4/8   & 86 \\
\midrule
Double layer     & 80  & 15   & -       & -  \\ \midrule
Double Layer v2     & 105 & 20  & -       & -  \\ \midrule
Triple     & 306 & 54  & -       & -  \\ \midrule
Quadruple & 367 & 65  & -       & -  \\
\midrule
Breast MNIST  & 546 & 78 & - & -  \\
 \bottomrule
\end{tabular}
\end{table}
\begin{figure*}
    \centering
\includegraphics[width=16cm]{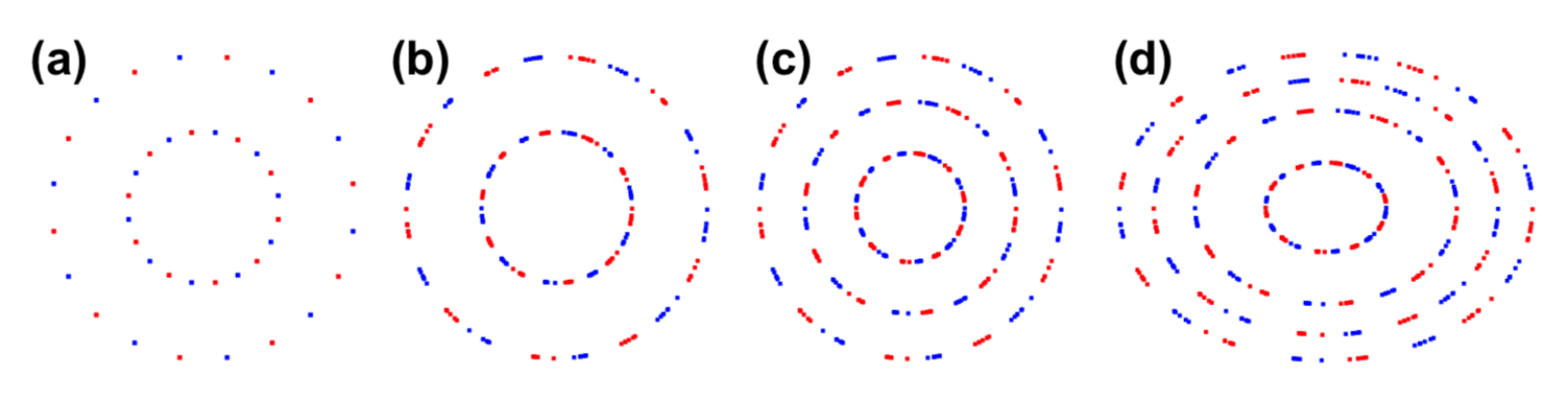}
    \caption{Visualization of the periodic data in different scenarios.}
    \label{fig_per_datasets}
\end{figure*}
\section{Experimental Results}\label{sec_III}
We have selected datasets that fall into two distinct categories, i.e., synthetic and real-world data. The synthetic datasets showcase various characteristics, including periodicity, noise, and feature correlations. The real-world dataset in our study is the MedMNIST \cite{Yang2023}, see the table \ref{dataset_details}. To ensure efficient computation of the Gram matrix, the number of samples for the training partitions of the datasets does not exceed $800$.


{\bf Classification Results:} To examine their performance, two objective functions are used for the classification problem; the $F1$ and the accuracy scores.
For the synthetic datasets, the $F1$ score is maximized in two different scenarios; the first one where the optimization process, only has access to the training and testing partitions of the data and in the second one, the optimization process maximizes the $F1$ score based on the cross-validation partition. That difference allows the optimization pipeline to produce over-fitted models in the first scenario, although the reported score is based on the testing set only.  

The table \ref{tab:exp1_results} shows the testing/training scores for each kernel  whereas table \ref{tab:exp2_results} summarizes the validation/testing/training scores, respectively.
By visually inspecting tables \ref{tab:exp1_results} and \ref{tab:exp2_results}, it is clear the performance of the proposed kernels, using the aforementioned two embedding techniques, are better than the RBF and the Squeezing kernels over different datasets. 
\begin{table*}[!hbt]
\centering
\caption{F$1$-scores (\%) across different Test/Train splits for first experimental scenario}
\label{tab:exp1_results}

\begin{tabular}{lcccccc}
\hline
\textbf{Dataset} &
  \textbf{KCS$^{-}$} &
  \textbf{KCS$^{+}$} &
  \textbf{Amp KCS$^{-}$} &
  \textbf{Amp KCS$^{+}$} &
  \textbf{RBF} &
  \textbf{Squeezing} \\ \hline
Moons v1 &
  \textbf{93.88}/93.33 &
  \textbf{93.88}/93.65 &
  \textbf{93.88}/93.33 &
  \textbf{93.88}/93.65 &
  \textbf{93.88/93.96} &
  \textbf{93.88}/93.33 \\ \hline
Moons v2 &
  \textbf{87.39}/96.56 &
  87.03/98.13 &
  87.39 / 96.56 &
  87.03/98.13 &
  86.24/\textbf{98.92} &
  87.16/97.98 \\ \hline
Circles v1 &
  84.21/82.59 &
  \textbf{85.11}/85.62 &
  84.21/82.59 &
  \textbf{85.11}/85.62 &
  83.67/84.0 &
  \textbf{85.11/86.38} \\ \hline
Circles v2 &
  \textbf{98.6}/96.16 &
  \textbf{98.6}/95.85 &
  \textbf{98.6}/96.16 &
  \textbf{98.6}/95.85 &
  \textbf{98.6}/\textbf{96.64} &
  \textbf{98.6}/96 \\ \hline
Hypercube v1 &
  \textbf{93.07}/98.04 &
  \textbf{93.07/99.53} &
  \textbf{93.07}/98.04 &
  \textbf{93.07/99.53} &
  91.43/91.8 &
  91.26/98.04 \\ \hline
Hypercube v2 &
  \textbf{87.39}/96.56 &
  66.46/\textbf{100.0} &
  \textbf{87.39}/96.56 &
  66.46/\textbf{100.0} &
  86.24/98.92 &
  87.16/97.98 \\ \hline
\end{tabular}
\end{table*}

\begin{table*}[!hbt]
\centering
\caption{F$1$-scores (\%) across different Cross validation/Test/Train splits for second experimental scenario.}
\label{tab:exp2_results}
\begin{tabular}{lcccccc}
\hline
\textbf{Dataset} &
  \textbf{KCS$^{-}$} &
  \textbf{KCS$^{+}$} &
  \textbf{Amp KCS$^{-}$} &
  \textbf{Amp KCS$^{+}$} &
  \textbf{RBF} &
  \textbf{Squeezing} \\ \hline
Moons v1 &
  \textbf{94.19}/\textbf{93.67}/94.88 &
  \textbf{94.19}/92.23/94.88 &
  \textbf{94.19}/93.17/94.88 &
  \textbf{95.19}/92.38/94.88 &
  \textbf{94.19}/94.24/91.67 &
  93.46/90.53/\textbf{94.92} \\ \hline
Moons v2 &
  93.83/94.88/94.14 &
  93.5/95.33/93.85 &
  93.83/94.88/94.14 &
  93.5/96.33/93.85 &
  96.69/96.68/94.14 &
  93.14/94.84/93.99 \\ \hline
Circles v1 &
  \textbf{84.04}/82.69/82.27 &
  83.38/82.35/\textbf{82.67} &
  \textbf{84.04}/82.69/82.27 &
  83.38/82.35/\textbf{82.67} &
  83.25/\textbf{83.22}/81.19 &
  83.0/80.0/82.27 \\ \hline
Circles v2 &
  97.68/97.63/97.36 &
  97.83/97.17/97.83 &
  97.68/97.63/97.36 &
  97.83/97.17/97.83 &
  96.21/95.65/97.36 &
  97.5/96.19/97.52 \\ \hline
Hypercube v1 &
  82.58/90.38/\textbf{93.81} &
  83.25/89.32/92.46 &
  82.58/90.38/\textbf{93.81} &
  83.25/89.32/92.46 &
  81.51/\textbf{100.0}/87.38 &
  \textbf{83.44}/89.32/93.11 \\ \hline
Hypercube v2 &
  82.79/82.19/91.14 &
  \textbf{84.28}/\textbf{84.4}/\textbf{96.57} &
  82.79/84.79/91.14 &
  \textbf{84.28}/\textbf{84.4}/\textbf{96.57} &
  82.36/81.31/82.41 &
  83.66/83.33/95.0 \\ \hline
\end{tabular}
\end{table*}

\begin{table*}[!hbt]
\centering
\caption{F$1$-scores (\%) for different test/train splits in the periodic datasets experiments.}
\label{tab:periodic_results}
\begin{tabular}{lccccccc}
\toprule
\textbf{Dataset} &
  \textbf{KCS$^-$} &
  \textbf{KCS$^+$} &
  \textbf{Amp KCS$^-$} &
  \textbf{Amp KCS$^+$} &
  \textbf{Squeezing} &
  \textbf{ESS} &
  \textbf{QEC} \\ \midrule
Disks v1  & 66.6 & 80.2 & 66.3 & 80.2 & 
42   & 90.2 & 90.2 \\ \midrule
Disks v2  & 100  & 100  & 99.2 & 99.3 
& 97   & 100  & 100  \\ \midrule
Triple    & 100  & 100  & 100  & 100  &
 98.4 & 100  & 100  \\ \midrule
Quadraple & 100  & 100  & 100  & 100  
& 98.3 & 100  & 100  \\ \bottomrule
\end{tabular}
\end{table*}
We consider the proposed kernels are benchmarked against a set of periodic datasets  as illustrated  in Fig.\ref{fig_per_datasets}, plots (a)-(d),  and compared to the ESS kernel.  Table.\ref{tab:periodic_results} reports testing $F1$ scores.  The proposed kernels outperform the Squeezing  kernel as seen before. However, the ESS and the associated quantum one, namely QEC are better at finding optimal decision boundaries.
In addition,  table.\ref{tab:deep-learning results}, shows the testing accuracy results of the proposed kernels against other kernels and also some deep learning neural networks that are considered as the state of the art for the same task. The proposed kernels outperform the deep-learning models with more than $1.3\%$.
\begin{table*}[!hbt]
\centering
\caption{Testing accuracy results on the Breast MNIST dataset compared with state-of-the-art models. The second row indicates testing accuracy (\%) results of the Breast MNIST dataset with $15\%$ noise added to the amplitude of the proposed kernels.}
\label{tab:deep-learning results}
\resizebox{\linewidth}{!}{%
\begin{tabular}{@{}cccccccccc@{}}
\toprule
\textbf{Dataset} &
  \textbf{KCS$^-$} &
  \textbf{KCS$^+$} &
  \textbf{Amp KCS$^-$} &
  \textbf{Amp KCS$^+$} &
  \textbf{RBF} &
  \textbf{Squeezing} &
  \textbf{Resnet-50 (224)} &
  \textbf{ResNet-18 (28)} &
  \textbf{auto-sklearn} \\ \midrule
Breast MNIST &
  86.5 &
  \textbf{86.67} &
  86.63 &
  \textbf{86.67} &
  79.0 &
  81.2 &
  84.2 \cite{Yang2023, Yang2021}&
  86.3 \cite{Yang2023, Yang2021}&
  80.3 \cite{Yang2023, Yang2021}\\ \bottomrule
  
Breast MNIST ($15\%$ noise) &
  81.5 &
  81.7 &
  \textbf{83.63} &
  82.67 &
  71.0 & 
  81.2 & 
  - & 
  - & 
  -
 \\ \midrule 
\end{tabular}}
\end{table*}

\begin{table*}[!hbt]
\centering
\caption{F$1$-score (\%) achieved with 10\% noise in the amplitude for noisy amplitude experiment. This result is from the Test/Train.}
\label{tab:exp2_results_noisy_10}
\begin{tabular}{lcccccc}
\hline
\textbf{Dataset} & \textbf{KCS$^{-}$} & \textbf{KCS$^+$} & \textbf{Amp KCS$^{-}$} & \textbf{Amp KCS$^+$} & \textbf{RBF} & \textbf{Squeezing} \\ \hline
Moons v1     & 90.81/91.2   & \textbf{92.28}/\textbf{91.5} & 90.32/91.3   & 90.88/91.35 & 88.3/90.46 & 90.9/91.33 \\ \hline
Moons v2     & 86.39/94.56  & 86.0/\textbf{95.3}  & 85.7/93.56 & \textbf{86.83}/94.13 & 84.24/92.1 & 85.6/93.98 \\ \hline
Circles v1   & 82.1/81.8    & \textbf{84.1}/\textbf{85.2}  & 82.2/81.9    & 83.1/84.62  & 80.67/81.0 & 81.11/81.3 \\ \hline
Circles v2   & 92.8/\textbf{95.73}   & \textbf{93.6}/94.0  & 93.3/95.6    & 93.5/94.85  & 92.6/94.14 & 93.1/94.3  \\ \hline
Hypercube v1 & 90.1/94.3    & 90.1/\textbf{96.0}    & \textbf{91.0}/93.04   & 89.07/93.1  & 88.3/89.93 & 89.8/93.3  \\ \hline
Hypercube v2 & 86.2/93.56 & 84.6/94.0  & 86.3/94.0    & \textbf{87.4}/\textbf{95.3}   & 82.24/92.9 & 84.6/92.8  \\ \hline
\end{tabular}
\end{table*}
To rigorously assess the performance of the proposed kernels in the optimization pipeline, random Gaussian noise has been introduced. This noise is specifically added to the amplitude of the KCS, with mean zero and standard deviations of $0.1$ and $0.15$. Additionally, noise is incorporated into the amplitude of the Squeezing kernel. For the Radial Basis Function (RBF), the noise is directly injected into the data. Table \ref{tab:exp2_results_noisy_10} showcases that the proposed kernels maintained their excellent performance compared to other kernels. Furthermore, the second row of Table \ref{tab:deep-learning results} reports the results of the Breast MNIST dataset with $15\%$ noise, indicating the robustness of the kernels against noise.
\begin{figure*}
\includegraphics[width=18cm]{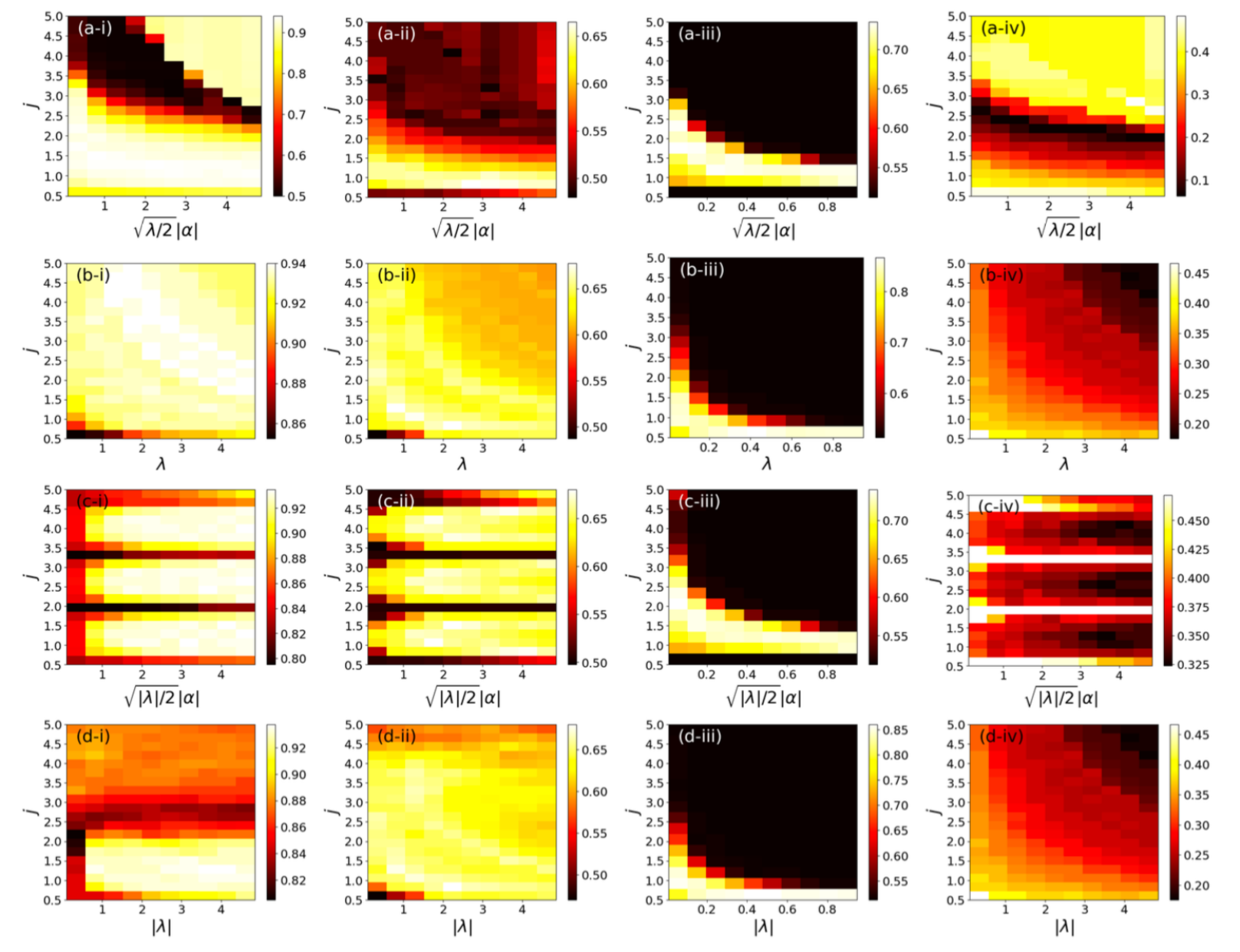}
\caption{Cross-validation analysis of kernels, including positive Kerr phase kernel, positive Kerr amplitude kernel, negative Kerr phase kernel, and negative Kerr amplitude kernel, is presented. Plots (a) to (d) depict the results obtained for different values of hyperparameters. Subplots (i) to (iv) illustrate the cross-validation outcomes for various data sets: Moons v2, Circles v2, Hypercube v1, and Double Layer v2.}\label{figur5_new} 
\end{figure*}


{\bf Analysing Hyperparameters:} 
We examine the influence of hyperparameters and their impact on the outcome. While the conventional machine learning scenario primarily focuses on fine-tuning the sole hyperparameter, namely $\gamma$, for optimal SVM results with the RBF kernel, the hyperparameters ${\lambda, j, \alpha}$ in the KCSs assume distinct roles.  

Figure \ref{figur5_new} provides comprehensive insights into the pivotal role of hyperparameters in determining SVM performance across various datasets: Moons v2, Circles v2, Hypercube v1, and Disks v2, depicted in plots (i)-(iv). 
More specifically, we investigate the role of discrete values of $j$ as both an integer and half-integer ranging from $0$ to $5$, along with selected discrete values of $\sqrt{\lambda/2}\alpha$ for the KCS$^\pm$ in plots (a) and (c) on the cross-validations. Plots (b) and (d) illustrate the influence of the hyper-parameters $ j $ and $\lambda$ within the same range for the amplitude KCS$^\pm$.   
The analysis underscores the significance of each parameter in influencing SVM performance. Notably, when comparing the roles of hyperparameters for KCS$^+$ and KCS$^-$, it becomes apparent that an increase in $j$ predominantly influences the performance. Conversely, in amplitude kernels, i.e., plots (b) and (d), the bandwidth $\lambda$ emerges as the dominant hyperparameter.
\section{Kernels implementation}\label{sec_IV}
The Kerr coherent states are categorized as a type of coherent states known as "non-linear coherent states." A non-linear coherent state is characterized by the utilization of a deformed displacement operator, wherein deformed annihilation and creation operators, denoted as $ \hat{A} = f(\hat{n})\hat{a} $ and $ \hat{A} = \hat{a}^{\dagger} f^{\dagger}(\hat{n}) $, respectively, substitute the conventional ones \cite{Dodonov2003}. A non-linear coherent state can be produced through the application of laser cooling and ion trapping techniques, as discussed in \cite{deMatosFilho1996}. When a trapped ion is subjected to laser fields, its internal and external degrees of freedom become interconnected through the interaction. Therefore, through careful adjustment of the laser fields, one can manipulate the quantized vibrational motion of the ion within the trap potential. The notable advantage of the exceedingly weak coupling between the vibrational modes and the external environment lies in its potential to facilitate the preparation and observation of non-linear coherent states with a remarkable level of stability, see appendix \cite{deMatosFilho1996,Vogel1995}.

Furthermore, we propose an approach for generating Kerr coherent states. This method involves the interaction of a two-level atom with a single-mode quantized cavity field, utilizing an intensity-dependent Jaynes–Cummings model, and concurrently exposing it to a strong external classical field. The temporal evolution of the system initially results in the formation of a superposition of Kerr coherent states. Depending on the initial states of the atom and the field, which can be suitably prepared, and considering the conditions under which the atom is detected (whether in the excited or ground state) following the interaction, the Kerr coherent state will collapse to the desired class of Kerr coherent states, see appendix \ref{sub_sec_cavity} for details.

Moreover, a proposed optical system facilitates the direct observation of a classical analog for the displacement of Fock states \cite{Perez-Leija2010}. This system involves a photonic lattice comprising evanescently coupled waveguides \cite{Christodoulides2003}, with a well-designed distribution of coupling between adjacent guides. Within these Glauber-Fock photonic lattices, each excited waveguide serves as a representation of a Fock state, and the spatial evolution of the light field corresponds to the probability amplitudes of the Kerr coherent state. Consequently, the emergence of these fundamental states and the underlying displacement process can be visually apprehended, see appendix \ref{sub_sec_photonic}.

\section{Discussion}
In this study, we have introduced the Kerr coherent states as a pivotal concept. We have elucidated their characterization as extensions of the $su(1,1)-$ and $su(2)-$coherent states, contingent upon the polarity of the Kerr parameter. Furthermore, we have underscored their intrinsic association with mathematical constructs inhabiting both the pseudo-sphere and sphere, exhibiting distinct curvatures that are contingent upon pertinent physical parameters, namely the frequency of light and the magnitude of the Kerr parameter.

By leveraging these coherent states as the foundation, we have formulated a methodology to map data, encompassing both phase encoding and amplitude encoding, into a structured feature space. Within this constructed feature space, a pseudo-Riemannian or Riemannian manifold is delineated, characterized by constant curvature. Potentially, this development offers a promising avenue for advancing quantum manifold optimization, particularly within the domain of hyperbolic machine learning, facilitated by photonic processors.

Furthermore, we have extended our investigation to encompass the application of the aforementioned Kerr Kernels, derived from Kerr coherent states, within the domain of Support Vector Machines (SVMs). It has been observed that through the utilization of phase encoding, the Kerr Kernels exhibit a broadened scope, resembling the characteristics of squeezed Kernels. Conversely, in the context of amplitude encoding, these kernels manifest as generalizations of cosine and hyperbolic cosine Kernels. Moreover, we have introduced novel kernels that correspond to the exponential sine squared function, a well-established kernel commonly employed in the analysis of periodic data.

We conducted an extensive analysis across multiple datasets to evaluate the efficacy of the Kerr kernels and to elucidate the influence of hyperparameters. Recognizing that noise can inherently accompany data encoded in both amplitude and phase, we deliberately introduced noise levels of up to $10\%$. Our findings demonstrate the robustness of the Kerr Kernels in the face of such perturbations, reaffirming their viability in real-world applications where data integrity may be compromised.


\section*{Acknowledgements}

The authors acknowledge the financial support by the Federal Ministry of Education and Research of Germany in the programme of “Souver"an. Digital. Vernetzt.”. Joint project 6G-life, project identification number: 16KISK002, and via grants 16KISQ039, 16KISQ077 and 16KISQ093. The further acknowledge financial support via the DFG Emmy-Noether program under grant number NO 1129/2-1.

%

%


\appendix

\section{Kerr Coherent states}\label{appenddix_A}
\begin{figure}[!h]
    \centering
    \includegraphics[width=9cm]{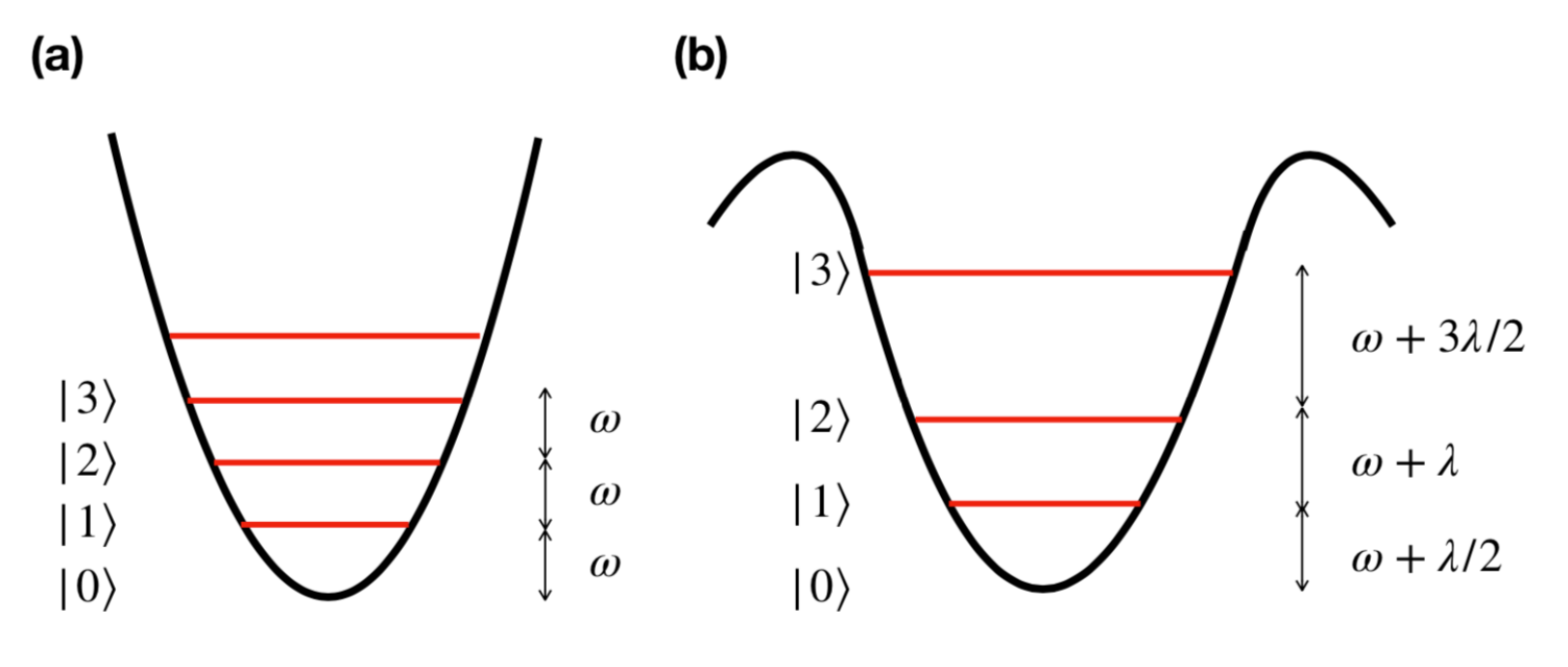}
    \caption{Schematically the spectrum of energy for a harmonic and anharmonic oscillator respectively in the plot (a) and (b). }
    \label{anharmonicity}
\end{figure}

A Kerr Hamiltonian can be defined as follows:
\begin{eqnarray}
    \hat{H}=\omega \hat{n} + \frac{\lambda}{2} \hat{n}^{2} 
\end{eqnarray}
in which $\hat{n}=\hat{a}^{\dagger}\hat{a}$ is the number operator and $\lambda$ is called Kerr parameter \cite{Scully1999}. The Hamiltonian represents an anharmonic oscillator in which the spacing between energy levels is anharmonic as schematically plotted in Fig. \ref{anharmonicity}. The Kerr Hamiltonian is widely used in different quantum systems, from  Josephson-Junction qubits, fiber optics, Kerr cavity, {\it etc.} 
\begin{definition}
    A Kerr coherent state is defined by applying the Kerr displacement operator
    \begin{eqnarray}
        D(\alpha)=\exp \left[\alpha \hat{A}^{\dagger}-\alpha^{\ast}\hat{A}\right]
    \end{eqnarray}
    on the reference state $\ket{0}$.
    For the positive Kerr parameter, the ladder operators are defined by 
    \begin{eqnarray}
    \hat{A}&=&\sqrt{\frac{\lambda}{2}}\hat{a} \sqrt{2j-1+\hat{n}}\nonumber\\
\hat{A}^{\dagger}&=&\sqrt{\frac{\lambda}{2}} \sqrt{2j-1+\hat{n}} \hat{a}^{\dagger}
    \end{eqnarray}
    For the negative Kerr parameter, the ladder operators are given by
    \begin{eqnarray}
    \hat{A}&=&\sqrt{\frac{|\lambda|}{2}}\hat{a} \sqrt{2j+1-\hat{n}}\nonumber\\  \hat{A}^{\dagger}&=&\sqrt{\frac{|\lambda|}{2}} \sqrt{2j+1-\hat{n}}\  \hat{a}^{\dagger}
    \end{eqnarray}     
\end{definition}
\begin{lemma}\label{lemma_decompos}
Let us define $[\hat{A},\hat{A}^{\dagger}]=2\hat{K}_{0}$. Then the operators $\hat A$, $\hat A^\dag$ and $\hat K_0$ satisfy the following commutation relations:
    \begin{eqnarray}\label{eq_comu}
[\hat{K}_{0},\hat{A}] =-\frac{\lambda}{2}\hat{A}, \ 
[\hat{K}_{0},\hat{A}^{\dagger}]=\frac{\lambda}{2}\hat{A}^{\dagger}
    \end{eqnarray}
    Then the Gaussian decomposition of  the  displacement operator $D(\alpha,\lambda)=\exp\left[\alpha \hat{A}^{\dagger}-\alpha^{\ast}\hat{A}\right]$ is given by
    \begin{eqnarray}\label{gbqca7}
    D(\alpha,\lambda)=e^{\alpha \hat A^{\dagger}-\alpha^{\ast}\hat A}=e^{\zeta \hat{A}^{\dagger}}e^{\ln[\zeta_{0}] \hat{K}_{0}} e^{-\zeta^{\ast}\hat{A}}
    \end{eqnarray}
    where in case that $\lambda>0$
    \begin{eqnarray}
    \zeta &=&\frac{\alpha}{|\alpha|}
    \sqrt{\frac{2}{\lambda}} \tanh\left(\sqrt{\frac{\lambda}{2}}|\alpha|\right),
    \label{def:zeta}
    \\
    \zeta_{0} &=& \cosh^{-j}\left(\sqrt{\frac{\lambda}{2}}|\alpha|\right)\label{eq16}
    \end{eqnarray}
    and in the case of negative Kerr parameter, we have 
    \begin{eqnarray}
    \zeta &=&\frac{\alpha}{|\alpha|}
    \sqrt{\frac{2}{|\lambda|}} \tan\left(\sqrt{\frac{|\lambda|}{2}}|\alpha|\right),
    \label{def:zeta-negative-lambda}
    \\
    \zeta_{0} &=& \cos^{j}\left(\sqrt{\frac{|\lambda|}{2}}|\alpha|\right)\label{eq16-negative-lambda}.
    \end{eqnarray}
    \begin{proof}
        Let us consider operator $F(t)$, with parameter $t$ defined by
    \begin{eqnarray}
    F(t)=e^{\left[\beta A^{\dagger}-\beta^{\ast}A\right]t}=e^{\zeta_{+}(t) A^{\dagger}}e^{\ln[\zeta_{0}(t)] K_{0}} e^{-\zeta_{-}(t) A}
    \end{eqnarray}
    where $\zeta_{\pm}(t)$ and $\zeta_{0}(t)$ are c-number functions of parameter $t$ to be determined under the following conditions:  $\zeta_{\pm}(0) = 0$ and $\zeta_{0}(0) = 1$. Once $\zeta_{\pm}(t)$ and  $\zeta_{0}(t)$ are determined, $\zeta_{\pm}$ and $\zeta_{0}$ are given by $\zeta_{\pm}(1)$ and $\zeta_{0}(1)$, respectively. \\
    Now we have
    \begin{eqnarray*}
    \frac{d}{dt}F&=&\left[\alpha A^{\dagger}-\alpha^{\ast}A\right] F\\
    &=&\dot{\zeta}_{+}\hat{A}^{\dagger}F\nonumber\\
    & &+e^{\zeta_{+}(t) A^{\dagger}}\frac{\dot{\zeta}_{0}}{\zeta_{0}}K_{0}e^{\ln[\zeta_{0}(t)] K_{0}} e^{-\zeta_{-}(t) A}\nonumber\\
    & &+e^{\zeta_{+}(t) A^{\dagger}}e^{\ln[\zeta_{0}(t)] K_{0}}\dot{\zeta}_{-}A e^{-\zeta_{-}(t) A}\\
    &=& \dot{\zeta}_{+}\hat{A}^{\dagger}F\nonumber\\
    &+& \frac{\dot{\zeta}_{0}}{\zeta_{0}}(K_{0}-\frac{\lambda}{2}\zeta_{+}A^{\dagger})F\label{eqgbqc44}\nonumber\\
    &-& \dot{\zeta}_{-} e^{-\lambda\ln[\zeta_{0}]/2}(A-2\zeta_{+}K_{0}+\frac{\zeta_{+}^{2}\lambda}{2}A^{\dagger})F\label{eqgbqc45}
    \end{eqnarray*}
Hence,  three  coupling differential equations are derived, i.e., 
\cite{Dehdashti2022}:
    \begin{eqnarray*}
    &&e^{-\lambda\ln[\zeta_{0}]/2}\frac{d\zeta_{-}}{dt}=\alpha^{\ast},\label{eqgbqc52}\\
    &&\zeta_{0}^{-1}\frac{d\zeta_{0}}{dt}+2e^{-\lambda\ln[\zeta_{0}]/2}\zeta_{+}\frac{d\zeta_{-}}{dt}=0\\
    &&\frac{d\zeta_{+}}{dt}-\frac{\lambda\zeta_{+}}{2\zeta_{0}}\frac{d\zeta_{0}}{dt}-e^{-\lambda\ln[\zeta_{0}]/2}\frac{\lambda\zeta_{+}^{2}}{2}\frac{d\zeta_{-}}{dt}=\alpha
    \end{eqnarray*}
    These can be reduced to
    \begin{eqnarray*}
        &&\zeta_{0}^{-1}\frac{d\zeta_{0}}{dt}+2\zeta_{+}\beta^{\ast}=0\\
        &&\frac{d\zeta_{+}}{dt}-\frac{\lambda\zeta_{+}}{2\zeta_{0}}\frac{d\zeta_{0}}{dt}-\frac{\lambda\zeta_{+}^{2}}{2}\beta^*=\beta
    \end{eqnarray*}
    The simple calculations gives the  differential equation: 
    \begin{eqnarray}
    \frac{d}{dt}\zeta_{+}+\frac{\lambda}{2}\beta^{\ast}\zeta_{+}^{2}=\beta
    \end{eqnarray}
    which, under the condition $\zeta_+(0)=0$ and $\lambda>0$, has the following solution:
    \begin{eqnarray}
    \zeta_{+}&=&\sqrt{\frac{\beta}{\beta^{\ast}}}  \sqrt{\frac{2}{\lambda}} \tanh\left[t\sqrt{\frac{\lambda}{2}}|\beta|\right]\nonumber\\
    &=&e^{i\phi}\sqrt{\frac{2}{\lambda}} \tanh\left[t\sqrt{\frac{\lambda}{2}}|\beta|\right]\label{eqgbqc57}
    \end{eqnarray}
    in which  $\beta=|\beta|e^{i\phi}$. Now by using the equation (\ref{eqgbqc52}), we write
    \begin{eqnarray}\label{eqgbqc58}
    \zeta_{0}=\cosh^{-4/\lambda} \left[t\sqrt{\frac{\lambda}{2}}|\beta|\right].
    \end{eqnarray}
    Now, by considering $t=1$ in the relations (\ref{eqgbqc57}) and (\ref{eqgbqc58}), the desired results are obtained.

    Note that Baker-Hausdorff lemma and the commutation relations (\ref{eq_comu}) can be used to achieve the following relations: 
    \begin{eqnarray*}
    e^{\zeta_{+}(t) A^{\dagger}}K_{0}e^{\zeta_{-}(t) A}
    =K_{0}-\frac{\lambda}{2}\zeta_{+}A^{\dagger}\\
    e^{\zeta_{+}(t) A^{\dagger}}e^{\ln[\zeta_{0}(t)] K_{0}}A e^{\ln[\zeta_{0}(t)] K_{0}}e^{\zeta_{-}(t) A}=\zeta_{0}^{-\lambda/2}\nonumber\\
    \times \left(A-2\zeta_{+}K_{0}+\frac{\zeta_{+}^{2}\lambda}{2}A^{\dagger}\right)
    \end{eqnarray*}

    In the case $\lambda<0$ then we can use the fact that $\cosh{ix}=\cos{x}$ and $\sinh{ix}=i\sin{x}$ to rewrite the relations (\ref{eqgbqc57}) and (\ref{eqgbqc58}) as follows:
    \begin{eqnarray}
    \zeta_{+}&=&\sqrt{\frac{\beta}{\beta^{\ast}}}  \sqrt{\frac{2}{|\lambda|}} \tan\left[t\sqrt{\frac{|\lambda|}{2}}|\beta|\right]\nonumber\\
    &=&e^{i\phi}\sqrt{\frac{2}{|\lambda|}} \tan\left[t\sqrt{\frac{|\lambda|}{2}}|\beta|\right]\label{eqgbqcp57}
    \end{eqnarray}
    and
    \begin{eqnarray}\label{eqgbqcp58}
    \zeta_{0}=\cos^{4/|\lambda|} \left[t\sqrt{\frac{|\lambda|}{2}}|\beta|\right].
    \end{eqnarray}
    \end{proof}
\end{lemma}
\begin{lemma}
    The displacement Kerr coherent states are given by
    \begin{eqnarray}  \ket{\alpha;\lambda^{+},j}&=&\cosh^{-2j}\left[\sqrt{\frac{\lambda}{2}} |\alpha| \right] \sum_{n=0}^{\infty} \sqrt{\frac{\Gamma(2j+n)}{\Gamma(2j) n!}} e^{-i n\phi}\nonumber \\
    &\times &\tanh^{n}\left[\sqrt{\frac{\lambda}{2}} |\alpha| \right] \ket{n}, \ \lambda>0 \label{eqa_coherent_pos}
    \end{eqnarray}
    and 
    \begin{eqnarray}  \ket{\alpha;\lambda^{-},j}&=&\cos^{2j}\left[\sqrt{\frac{|\lambda|}{2}} |\alpha| \right] \sum_{n=0}^{2j} \sqrt{\frac{(2j)!}{(2j-n)!n!}} e^{-i n\phi}\nonumber \\
    &\times &\tan^{n}\left[\sqrt{\frac{|\lambda|}{2}} |\alpha| \right] \ket{n}, \ \lambda<0 \label{eqa_coherent_neg}
    \end{eqnarray}
    \begin{proof}
        Using the Gaussian decomposition introduced in the lemma \ref{lemma_decompos} gives the result directly.
    \end{proof}
\end{lemma}
\begin{prop}
    The Kerr coherent states given by equations (\ref{eqa_coherent_pos}) and (\ref{eqa_coherent_neg}) correspondingly simplify to the $su(1,1)$ and $su(2)$ coherent states when the values of $\lambda$ are set to $2$ and $-2$, respectively.
\end{prop}
The approach used to explore the structure of the Kerr coherent state manifolds involves examining their Fubini-Study metric, which is defined by the infinitesimal distance $ds$ between two
neighboring Kerr coherent states.
\begin{definition}
    A Fubini-Study metric is defined by considering the differential distance $ds$ between two
neighboring pure quantum states $\ket{\psi(x^{\mu})}$ and $\ket{\psi(x^{\mu}+dx^{\mu})}$:
\begin{eqnarray}
    ds^{2}=g_{\mu\nu}dx^{\mu}dx^{\nu}
\end{eqnarray}
where 
\begin{eqnarray}\label{eq_meric_co}
g_{\mu\nu}=\frac{\braket{\psi_{\mu}}{\psi_{\nu}}}{\braket{\psi}{\psi}}-\frac{\braket{\psi_{\mu}}{\psi}\braket{\psi}{\psi_{\nu}}}{|\braket{\psi}{\psi}|^{2}}
\end{eqnarray}
in which $x^{\mu}$
is a set of real parameters which define the state  $\ket{\psi}$ and $\ket{\psi_{\mu}}=\frac{\partial}{\partial x^{\mu}}\ket{\psi}$.
\end{definition}
\begin{lemma}
    The Fubini-Study metrics of the Kerr coherent state, for the positive and negative Kerr parameter, are respectively given by
    \begin{eqnarray}
        ds^{2}&=&j\lambda dr^{2}+\frac{j}{2}\sinh^{2}\left[\sqrt{2\lambda}r\right]d\phi^{2}\label{metric_pos}\\
        ds^{2}&=&j\lambda dr^{2}+\frac{j}{2}\sin^{2}\left[\sqrt{2\lambda}r\right]d\phi^{2}\label{metric_neg}
         \end{eqnarray}
   while the following stands for the coherent state one: 
    \begin{eqnarray}
        ds^{2}= dr^{2}+r^{2}d\phi^{2}
    \end{eqnarray}
    \begin{proof}
        Let us define the non-normalized Kerr coherent state as
        \begin{eqnarray}       \ket{z}=\sum_{n=0}^{\infty}\sqrt{\frac{\Gamma(2j+n)}{\Gamma(2j)n!}} z^{n}\ket{n},
        \end{eqnarray}
        where $z=\tanh\left[\sqrt{\frac{\lambda}{2}r}\right]e^{-i\phi}$. By replacing the following relations
        \begin{eqnarray}
            \braket{z}{z}&=&(1-|z|^{2})^{-2j}\nonumber\\
            \braket{z}{z_{z}}&=& 2j \bar{z}(1-|z|^{2})^{-2j-1}dz\nonumber\\
            \braket{z_{z}}{z_{z}}&=& 2j (1-|z|^{2})^{-2j-2}(1+2j|z|^{2})d\bar{z}dz\nonumber
        \end{eqnarray}
        in the equation (\ref{eq_meric_co}), we obtain:
        \begin{eqnarray}
            ds^{2}=2j\frac{dzd\Bar{z}}{(1-|z|^{2})^{2}}
        \end{eqnarray}
        Now, by considering the fact that 
        \begin{eqnarray}\nonumber
        dz=e^{-i\phi}\cosh^{-2}\left[\sqrt{\frac{\lambda}{2}}r\right]\left(\sqrt{\frac{\lambda}{2}}dr-\frac{\sinh\left[\sqrt{2\lambda}r\right]}{2}d\phi\right),
        \end{eqnarray}
         the above metric can be rewritten as
         \begin{eqnarray}
             ds^{2}=j\lambda dr^{2}+\frac{j}{2}\sinh^{2}\left[\sqrt{2\lambda}r\right]d\phi^{2}
         \end{eqnarray}
\indent In the case of negative value of the Kerr parameters,  the non-normalized coherent state is defined as
    \begin{eqnarray}
        \ket{z}=\sum_{n=0}^{2j} \sqrt{\frac{(2j)!}{(2j-n)!n!}}z^{n} \ket{n}
    \end{eqnarray}
    in which $z=e^{-i n\phi}
\tan\left[\sqrt{\frac{|\lambda|}{2}} |\alpha| \right]$. Following the above method leads into the relations:
\begin{eqnarray}
            \braket{z}{z}&=&(1+|z|^{2})^{2j}\nonumber\\
            \braket{z}{z_{z}}&=& 2j \bar{z}(1+|z|^{2})^{2j-1}dz\nonumber\\
            \braket{z_{z}}{z_{z}}&=& 2j (1+|z|^{2})^{2j-2}(1+2j|z|^{2})d\bar{z}dz\nonumber
        \end{eqnarray}
        which gives the Fubini-Study metric,
        \begin{eqnarray}
        ds^{2}=\frac{2jdzd\Bar{z}}{(1+|z|^{2})^{2}}
        \end{eqnarray}
        \indent In the case of normal coherent states, the non-normilzed coherent state is defined by $\ket{z}=\sum_{n=0}^{\infty}z^{n}/n! \ket{n}$, in which $z=re^{i\phi}$. Using the following relations
    \begin{eqnarray*}
        \braket{z}{z}&=&e^{|z|^{2}}\\
        \braket{z}{z_{z}}&=&e^{|z|^{2}}\bar{z}dz\\
        \braket{z_{z}}{z_{z}}&=&(1+|z|^{2})e^{|z|^{2}}d\bar{z}dz,
    \end{eqnarray*}
    leads to the Fubini-Study metric as $ds^{2}=dzd\Bar{z}$.  
    \end{proof}   
\end{lemma}
\begin{lemma}
The feature space derived from encoding data into Kerr coherent states is a two-dimensional manifold with uniform curvature. Specifically, the feature space created by utilizing positive Kerr coherent states forms a Hyperbolic plane $\mathbb{H}$, whereas the feature space constructed with negative Kerr coherent states constitutes a spherical plane $\mathbb{S}$.
\begin{proof}
    The non-zero Christoffel symbols, defined as $2\Gamma_{jk}^{i}=g^{il}\left(\partial_{k}g_{lj}+\partial_{j}g_{lk}-\partial_{l}g_{jk}\right)$, where $g^{il}$ is the inverse metric $g_{il}$ and $\partial_{i}=\frac{\partial}{\partial x_{i}}$, of the Fubini-Study metric are given by 
\begin{eqnarray*}
\Gamma^{\phi}_{r\phi}&=&\sqrt{2\lambda}\coth\left[\sqrt{2\lambda } r\right], \\ \Gamma^{r}_{\phi\phi}&=&-\sqrt{\frac{\lambda}{2}} \sinh\left[2\sqrt{2\lambda } r\right]
\end{eqnarray*}
The non-zero element of the Riemann tensor defined 
\begin{eqnarray}
    R^{l}_{ijk} =\partial_{i}\Gamma_{jk}^{l}-\partial_{j}\Gamma_{ik}^{l}-\Gamma_{jm}^{l}\Gamma_{ik}^{m}+\Gamma_{im}^{l}\Gamma_{jk}^{m}
\end{eqnarray}
is obtained by 
\begin{eqnarray}
    R^{r}_{r\phi r \phi}=-2\lambda \sinh^{2}\left[\sqrt{2\lambda } r\right] 
\end{eqnarray}
 We find the non-zero Ricci tensor as the following:
\begin{eqnarray}
    R_{rr}= -4\lambda^{2},\
    R_{\phi \phi} =-2\lambda \sinh^{2}\left[\sqrt{2\lambda } r\right]
\end{eqnarray}
which leads to the constant Ricci scalar, i.e.,
\begin{eqnarray}
    R=g^{rr}R_{rr}+g^{\phi\phi}R_{\phi\phi}=-\frac{2\lambda}{j}
\end{eqnarray}
which indicates the metric has the negative constant.    In addition, we can define the following parametrical equations as follows:   
    \begin{eqnarray}
    x_{0}&=&\sqrt{\frac{j}{2}}\cosh\left[\sqrt{2\lambda}r\right]\\
    x_{1}&=&\sqrt{\frac{j}{2}}\sinh\left[\sqrt{2\lambda}r\right] \cos{\phi}\\        x_{2}&=&\sqrt{\frac{j}{2}}\sinh\left[\sqrt{2\lambda}r\right] \sin{\phi}
    \end{eqnarray}
    which is defined as a conformal Minkowski space, 
    \begin{eqnarray}
        ds^{2}=dx_{0}^{2}-dx_{1}^{2}-dx_{2}^{2}
    \end{eqnarray}
    and defines the following hyperbolic sheet:
    \begin{eqnarray}
        x_{0}^{2}-x_{1}^{2}-x_{2}^{2}=\frac{j^{2}}{4}
    \end{eqnarray}
For the metric (\ref{metric_neg}), the  non-zero Christoffel symbols are given by
\begin{eqnarray*}
\Gamma^{\phi}_{r\phi}&=&\sqrt{2\lambda}\cot\left[\sqrt{2\lambda } r\right], \\ \Gamma^{r}_{\phi\phi}&=&-\sqrt{\frac{\lambda}{2}} \sin\left[2\sqrt{2\lambda } r\right]
\end{eqnarray*}
which leads to the following Ricci scalar
\begin{eqnarray}
    R=\frac{2|\lambda|}{j}
\end{eqnarray}
which indicates the positive constant curvature. Moreover, we  define the following parametrical equa-
tions as follows
    \begin{eqnarray}
    x_{0}&=&\sqrt{\frac{j}{2}}\cos\left[\sqrt{2\lambda}r\right]\\
    x_{1}&=&\sqrt{\frac{j}{2}}\sin\left[\sqrt{2\lambda}r\right] \cos{\phi}\\        x_{2}&=&\sqrt{\frac{j}{2}}\sin\left[\sqrt{2\lambda}r\right] \sin{\phi}
    \end{eqnarray}
which gives a spherical space embedded in the three dimension, i.e.,
\begin{eqnarray}
x_{0}^{2}+x_{1}^{2}+x_{2}^{2}=\frac{j}{2}
\end{eqnarray}
\end{proof}
\end{lemma}
\begin{lemma}\label{res_iden}
    For each fixed $\lambda$, the Kerr coherent states  give rise to a resolution of unity,  
    \begin{eqnarray}\label{eq_resolution}
        \int d\mu \ket{\alpha,\lambda}\bra{\alpha,\lambda}=\mathbbm{1} 
    \end{eqnarray}
    with the corresponding invariant measure i.e.,
    \begin{eqnarray}
        d\mu&=&\frac{2j+1}{2\pi}\sqrt{\frac{|\lambda|}{2}}\sin\left[\sqrt{2|\lambda|}r\right]drd\phi, \lambda<0,\label{meas_neg}\\
        d\mu&=&\frac{2j-1}{2\pi}\sqrt{\frac{\lambda}{2}}\sinh\left[\sqrt{2\lambda}r\right]drd\phi, \lambda>0,\label{meas_pos}\\
    \end{eqnarray} 
    \begin{proof}
     By substituting the relation (\ref{eqa_coherent_neg}) into the right side of the above equation  the above equation and the following identity
     \begin{eqnarray}
         \delta_{n,m}=\frac{1}{2\pi}\int_{0}^{2\pi}d\phi e^{-i\phi(n-m)}
     \end{eqnarray}
     where $\delta_{n,m}$ denotes Kronecker-Delta, we have 
     \begin{eqnarray*}
     &&\sqrt{2\lambda}\sum_{n=0}^{2j}\frac{(2j+1)!}{(2j-n)!n!}
    \int_{0}^{\frac{\pi}{\sqrt{2\lambda}}}dr \sin\left[\sqrt{2|\lambda|}r\right]\\ 
    &\times &\cos^{4j}\left[\sqrt{\frac{\lambda}{2}}r\right] \tan^{2n}\left[\sqrt{\frac{\lambda}{2}}r\right]\ketbra{n}{n}\\
    &=&\sum_{n=0}^{2j}\frac{(2j+1)!}{(2j-n)!n!}
    \int_{0}^{\pi}d(\cos{\vartheta})\left(\frac{1+\cos{\vartheta}}{2}\right)^{2j-n} \\
    &\times & \left(\frac{1-\cos{\vartheta}}{2}\right)^{n}\ketbra{n}{n}\\
&=& \sum_{n=0}^{2j}\frac{(2j+1)!}{(2j-n)!n!} \int_{0}^{1}dx x^{2j-n}(1-x)^{n}\ketbra{n}{n}\\
&=&\sum_{n=0}^{2j}\ketbra{n}{n}=\mathbbm{1}.
     \end{eqnarray*}
     In the last step of the proof, we applied:
     \begin{eqnarray}
         \int_{0}^{1}dx x^{2j-n}(1-x)^{n}=\frac{(2j-n)!n!}{(2j+1)!}
     \end{eqnarray}
     By substituting the relation (\ref{eqa_coherent_pos}) into the right side of the  relation (\ref{eq_resolution}), we obtain 
     \begin{eqnarray*}
     && \sqrt{2\lambda}\sum_{n=0}^{\infty}\frac{\Gamma(2j+n)}{\Gamma(2j-1)n!}\int_{0}^{\infty}dr \sinh\left[\sqrt{2|\lambda|}r\right]\\ 
    &\times &\cosh^{-4j}\left[\sqrt{\frac{\lambda}{2}}r\right] \tanh^{2n}\left[\sqrt{\frac{\lambda}{2}}r\right]\ketbra{n}{n}\\
    &=&\sum_{n=0}^{\infty}\frac{\Gamma(2j+n)}{\Gamma(2j-1)!n!}
    \int_{1}^{\infty}d(\cosh{\vartheta})\left(\frac{\cosh{\vartheta}+1}{2}\right)^{-2j-n} \\
    &\times & \left(\frac{\cosh{\vartheta}-1}{2}\right)^{n}\ketbra{n}{n}\\
&=& \sum_{n=0}^{2j}\frac{\Gamma(2j+n)}{\Gamma(2j-1)n!} \int_{1}^{\infty}dx x^{-2j-n}(1-x)^{n}\ketbra{n}{n}\\
&=&\sum_{n=0}^{2j}\ketbra{n}{n}=\mathbbm{1}.
     \end{eqnarray*}
     in which the following identity is applied:
     \begin{eqnarray}
         \int_{1}^{\infty}dx x^{-2j-n}(1-x)^{n}=\frac{\Gamma(2j-1)n!}{\Gamma(2j+n)}
     \end{eqnarray}
    \end{proof}
\end{lemma}
\begin{definition}
    A Kerr kernel can be defined as the overlap of two Kerr coherent states:
    \begin{eqnarray}
        K(\alpha_{1},\alpha_{2})=\braket{\alpha;\lambda}{\alpha;\lambda}
    \end{eqnarray}
  where direct calculation gives the following result for the positive Kerr parameter \begin{eqnarray}\label{ap_eq_inner_pos}
   K_{\lambda^{+}}(\alpha_{1},\alpha_{2})=
\frac{\left(\sech^{2}\left[\sqrt{\frac{\lambda}{2}}r_{1} \right]\sech^{2}\left[\sqrt{\frac{\lambda}{2}}r_{2}\right]\right)^{j}}{\left[1-e^{i(\theta_{1}-\theta_{2})}\tanh\left(\sqrt{\frac{\lambda}{2}}r_{1}\right)\tanh\left(\sqrt{\frac{\lambda}{2}}r_{2}\right)\right]^{2j}}.\nonumber\\ 
\end{eqnarray}
and for $\lambda \in \mathbb{R}^{-}$, is given by
\begin{eqnarray}\label{ap_eq_inner_neg}
    K_{\lambda^{-}}(\alpha_{1},\alpha_{2})=\frac{\left[1+e^{i(\theta_{1}-\theta_{2})}\tan\left[\sqrt{\frac{|\lambda|}{2}}r_{1}\right]\tan\left[\sqrt{\frac{|\lambda|}{2}}r_{2}\right]\right]^{2j}}{\left(\sec^{2}\left[\sqrt{\frac{|\lambda|}{2}}r_{1}\right]\sec^{2}\left[\sqrt{\frac{|\lambda|}{2}}r_{2}\right]\right)^{j}}.\nonumber\\
\end{eqnarray}  
\end{definition}
\begin{lemma}
    The Kerr kernels exhibit the property of reproducibility, i.e.,
    \begin{eqnarray}
    K(\alpha_{1},\alpha_{2})=\int d\mu(\alpha) K(\alpha_{1},\alpha) K(\alpha,\alpha_{2})
    \end{eqnarray}
    \begin{proof}
       Using \ref{res_iden} leads to the result directly. 
    \end{proof}
\end{lemma}
\section{Kerr Coherent States' Implementation}\label{sec.app.b}
In the following, we develop two methods to implement Kerr coherent states. The first method involves a quantum optics setup, while the second method is based on simulating the Kerr coherent states in a set of coupled waveguides.

\subsection{Generation of Kerr coherent states}\label{sub_sec_cavity}
In our analysis, we examine a Hamiltonian that describes the interaction between a two-level atom and a single-mode cavity field. This interaction is subject to an intensity-dependent coupling, along with the influence of an external classical field. By taking  the account the resonance conditions and apply the rotating-wave approximation, the Hamiltonian can be expressed in the following \cite{Solano2003,Zou2004,Miry2012}:
\begin{eqnarray}\label{eq_Ham}
    \hat{H}&=&g \left(e^{i\varphi}\sigma_{-}+e^{-i\varphi}\sigma_{+}\right)\nonumber\\
    &+&\Omega \left(\sigma_{-}\hat{A}^{\dagger}+\sigma_{+}\hat{A}\right)
\end{eqnarray}
where parameters $g$ and $\Omega$ 
 are the coupling coefficients of the atom
with classical and quantized cavity fields, $\sigma_{\pm}$ is the ladder operator, i.e., $\sigma_{-}=\ketbra{e}{g}$ and $\sigma_{+}=\ketbra{g}{e}$ in which $\ket{e}$ and $\ket{g}$ are respectively ground and excited states of the atom; $\hat{A}$ and $\hat{A}^{\dagger}$ are the Kerr annihilation and creation operators and $\varphi$  is the phase of classical
field. By using the following transformation:
\begin{eqnarray}\label{eq_r}
    R=\exp\left[\frac{\pi}{4}(\sigma_{+}-\sigma_{-})\right]\exp\left[\frac{i\varphi}{2}\sigma_{z}\right]
\end{eqnarray}
by which 
\begin{eqnarray}
    R \sigma_{\pm} R^{\dagger}=e^{\pm i \varphi}{2}\left[\sigma_{z}\pm (\sigma_{+}-\sigma_{-})\right]
\end{eqnarray}
the Hamiltonian (\ref{eq_Ham}) becomes
\begin{eqnarray}
H_{tran}&=&RHR^{\dagger}=g\sigma_{z}\\
    &+& \frac{\Omega}{2}\Big[ e^{-i\varphi}\hat{A}^{\dagger}\left(\sigma_{z}-(\sigma_{+}-\sigma_{-})\right)\nonumber\\
    &+& e^{i\varphi}\hat{A}\left(\sigma_{z}+(\sigma_{+}-\sigma_{-})\right) \Big]\nonumber
\end{eqnarray}
Hence, in the interaction picture, by using the unitary operator $\hat{T}(t)=\exp[ig\sigma_{z}t]$ the time evolution of $H_{tran}$ is given by
\begin{eqnarray}
H_{int}&=&e^{ig\sigma_{z}t}H_{tran}e^{-ig\sigma_{z}t}\\
    &=&\frac{\Omega}{2}\Big[e^{-i\varphi}\hat{A}^{\dagger}\left(\sigma_{z}-\left(\sigma_{+}e^{2igt}-\sigma_{-}e^{-2igt}\right)\right)\nonumber\\
&+&e^{i\varphi}\hat{A}\left(\sigma_{z}+\left(\sigma_{+}e^{2igt}-\sigma_{-}e^{-2igt}\right)\right)\Big]\nonumber
\end{eqnarray}
In the strong classical field regime, i.e., $g\ll \Omega$,   $e$ the term that oscillates
with high frequencies in the Hamiltonian is eliminated
\begin{eqnarray}\label{eq_H_eff}
    H_{eff}=\frac{\Omega}{2}\left[e^{-i\varphi}\hat{A}^{\dagger}+e^{i\varphi}\hat{A}\right]\sigma_{z}
\end{eqnarray}
The time evolution operator of Hamiltonian (\ref{eq_H_eff}) is given by
\begin{eqnarray}
    U_{eff}=\exp\left[\frac{-i\Omega t}{2}\left[e^{-i\varphi}\hat{A}^{\dagger}+e^{i\varphi}\hat{A}\right]\sigma_{z}\right]
\end{eqnarray}
which is  treated as a Kerr displacement
operator. 
We are able to come back to the standard time evolution, i.e., the time evolution is corresponded to the  the Hamiltonian (\ref{eq_Ham}), by using following transformations:
\begin{eqnarray}
U(t)&=&\hat{R}^{\dagger}\hat{T}^{\dagger}(t)\hat{U}_{eff}\hat{T}(0)\hat{R}\\
    &=&\exp\left[-igt\left(\sigma_{-}e^{i\varphi}+\sigma_{+}e^{-i\varphi}\right)\right]\nonumber\\
    &\times&
    \exp\left[\frac{-i\Omega t}{2}\left(e^{-i\varphi}\hat{A}^{\dagger}+e^{i\varphi}\hat{A}\right)\left(\sigma_{-}e^{i\varphi}+\sigma_{+}e^{-i\varphi}\right)\right]\nonumber
\end{eqnarray}
Now, by  assuming that the cavity is initially
prepared in the vacuum of the field and the atoms are in a
superposition of excited and ground states, i.e.,
\begin{eqnarray}
    \ket{\psi}=\frac{1}{\sqrt{2}}\left(\ket{g}+\ket{e}\right)\ket{0}
\end{eqnarray}
Using relation (\ref{eq_r}), we have 
\begin{eqnarray}
    \hat{R}\ket{\psi}=\left[\cos{\frac{\varphi}{2}}\ket{e}-i\sin{\frac{\varphi}{2}}\ket{g}\right]\ket{0}
\end{eqnarray}
Hence, the simple calculations leads into the final result:
\begin{eqnarray}
    \ket{\psi(t)}&=&\frac{1}{\sqrt{2}}\Big[e^{-i\varphi/2}e^{-igt}\cos{\frac{\varphi}{2}}\ket{\alpha;\lambda,j}\\
    &+&i e^{-i\varphi/2}e^{igt}\sin{\frac{\varphi}{2}}\ket{-\alpha;\lambda,j}
    \Big]\ket{e}\nonumber\\
    &+& \frac{1}{\sqrt{2}}\Big[e^{-i\varphi/2}e^{-igt}\cos{\frac{\varphi}{2}}\ket{\alpha;\lambda,j}\\
    &-&i e^{-i\varphi/2}e^{igt}\sin{\frac{\varphi}{2}}\ket{-\alpha;\lambda,j}
    \Big]\ket{g}
\end{eqnarray}
in which $\alpha=\frac{-i\Omega t}{2}e^{-i\varphi}$. Therefore, by setting $\varphi=0$, and measuring the atom state, the Kerr coherent state is obtained.
\subsection{ Photonic
Lattice Realization of Kerr Coherent States}\label{sub_sec_photonic}
\begin{figure}[t]
\includegraphics[width=9cm]{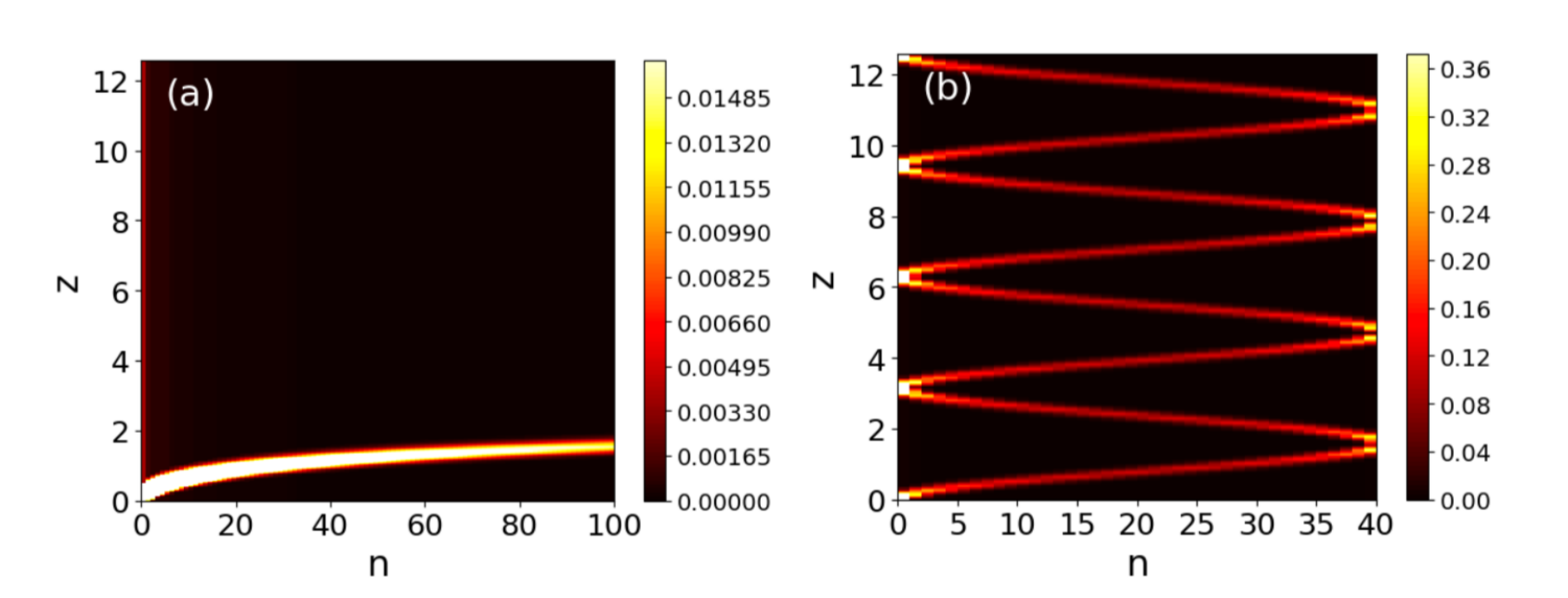}
\caption{The intensity evolution in a waveguide lattice for $|\lambda|=2$ and $j=20$ during n waveguides is depicted in plots (a) and (b) for the positive and negative Kerr coherent states, respectively.}
    \label{fig_psi_12}
\end{figure}
Optics typically involves the study of electromagnetic waves as continuous functions of space and time. However, there are instances where the evolution of an optical field can be treated as a discrete problem. A notable example of this is found in the study of light in a coupled 1D waveguide array. In such an array, numerous (potentially infinite) single-mode channel waveguides are arranged side by side, causing their individual modes to overlap. This arrangement leads to coupling between each waveguide and its nearest neighbors, significantly influencing the propagation of light in these waveguide arrays. 
In this scenario, the evolution of the optical mode electric field amplitude, denoted as $E_{n}$, at the n$^{\text{th}}$ site of the array, follows a discrete linear Schr\"{o}dinger-like
equation is given by:
\begin{eqnarray}\label{eq_Sch}
    i\frac{dE_{n}}{dz}+C_{n}E_{n-1}+C_{n+1}E_{n+1}=0
\end{eqnarray}
where in the weak coupling regime, $C_{n}$ depends exponentially
on the distance between the guides, i.e., $C_{n}=C_{1}\exp\left[-(d_{n}-d_{0}/\kappa)\right]$, in which $d_{1}$ and $\kappa$ are fitting parameter \cite{Keil2011,shahram2,Christodoulides2003}.
By choosing the following distance between the first site and n$^{\text{th}}$ site:
\begin{eqnarray}
    d_{n}=d_{0}-\kappa \ln \left[\frac{|\lambda|\sqrt{2j\mp1 \pm n}}{2C_{0}}\right]
\end{eqnarray}
the coefficient $C_{n}$ is given by
\begin{eqnarray}
    C_{n}=\frac{|\lambda|}{2}\sqrt{2j\mp 1 \pm n}
\end{eqnarray}

Now, we can write the differential equation (\ref{eq_Sch}) as the follows:
\begin{eqnarray}
    i\frac{d\Psi}{dz}=-(\hat{A}+\hat{A}^{\dagger})\Psi
\end{eqnarray}
where $\Psi=\sum_{n=0}^{s}E_{n}\ket{n}$. Therefore, we have 
\begin{eqnarray} \Psi(z)=e^{i(\hat{A}+\hat{A}^{\dagger})z}\ket{0}
\end{eqnarray}
the sign $\pm$ implies the sign of the Kerr parameter. Therefore, By selecting specific values for $z$ corresponding to each data point, we can effectively encode the data into the photonic lattice.
Fig \ref{fig_psi_12}, plots (a) and (b), illustrate light propagation in the waveguide lattice for the positive and negative Kerr parameters.
\end{document}